\documentclass[aps,prl,preprint,superscriptaddress,groupedaddress]{revtex4-1}

\usepackage{graphicx,amsmath,bm}
\usepackage[caption=false]{subfig}
\usepackage{color}
\usepackage{natbib}

\begin{document}

\title{Short Electron Bunch Generation Using Single-Cycle Ultrafast Electron Guns}

\author{{ Arya Fallahi$^{1,*}$, Moein Fakhari$^1$, Alireza Yahaghi$^1$, Miguel Arrieta$^1$, and Franz X. K\"artner$^{1,2,3}$} \\
{\small \em $^1$ Center for Free-Electron Laser Science, DESY, Notkestrasse 85, 22607 Hamburg, Germany \\
$^2$ Department of Physics, University of Hamburg, Jungiusstrasse 9, 20355 Hamburg, Germany\\
$^3$ The Hamburg Center for Ultrafast Imaging, Luruper Chaussee 149, 22761 Hamburg, Germany }}

\date{\today}

\begin{abstract}
We introduce a solution for producing ultrashort ($\sim$fs) high charge ($\sim$pC) from ultra-compact guns utilizing single-cycle THz pulses.
We show that the readily available THz pulses with energies as low as 20$\,\mu$J are sufficient to generate multi-10\,keV electron bunches.
Moreover, It is demonstrated that THz energies of 2\,mJ are sufficient to generate relativistic electron bunches with higher than 2\,MeV energy.
The high acceleration gradients possible in the structures provide 30\,fs electron bunches at 30\,keV energy and 45\,fs bunches at 2\,MeV energy.
These structures will underpin future devices for strong field THz physics in general and miniaturized electron guns, in which the high fields combined with the short pulse duration enable electron beams with ultrahigh brightness.
\end{abstract}

\maketitle

The achievable acceleration gradient in an accelerator device is known to be the main limiting factor governing the emittance and consequently the length of the output bunch.
In a conventional particle accelerator, the electrical breakdown of metals introduces a strong limitation on the accelerating fields which are typically 10-100\,MV/m \cite{Loew1988}.
This fact turns out to be the major limit determining the maximum accelerating gradients in many large scale facilities like SLAC \cite{Wang1985}, CERN's compact linear collider (CLIC) \cite{linssen2012physics} and the design of the next linear collider (NLC) \cite{Phinney2002}.
Moreover, the small acceleration gradient dictates long accelerator lengths, making it the main impediment in developing compact and therefore lower cost devices based on beams of particles with relativistic energies.
The desire to realize compact accelerators has spurred much research into the use of alternative acceleration schemes, such as dielectric laser acceleration (DLA) \cite{Peralta2013,England2014,Breuer2013}, laser-driven plasma acceleration (LPA) \cite{tajima1979,Malka2002,Leemans2006,wang2013,leemans2014,steinke2016}, and THz acceleration \cite{Nanni2015,Wong2013,Yoder2005}. \footnotetext[1]{Corresponding authors: Arya Fallahi (arya.fallahi@cfel.de)}

The empirical studies done by Loew and Wang \cite{Loew1988,Wang1989} had initially shown that electron field emission, scaling as $f^{1/2}/\tau^{1/4}$ with $f$ the operation frequency, and $\tau$ the pulse duration of the accelerating field, is the main reason for electrical breakdown \cite{Kilpatrick1957}.
The above approximate scaling behavior justified research towards higher operating frequencies and ultrafast schemes to achieve compact accelerators \cite{Braun2003}.
However, the recent comprehensive study on breakdown thresholds of various accelerators by Laurent et al. \cite{Laurent2011} and Dolgashev et al. \cite{Dolgashev2010} demonstrated that pulsed heating of the accelerator walls is the dominant factor limiting acceleration gradients.
This conclusion confirmed the observed lower operational gradients in existing facilities when compared with predictions from the previously derived scaling laws.
The authors concluded that the pulse duration of the accelerating field plays the major role in the breakdown event, since it is directly linked to the pulse energy governing the pulsed heating in the device.
Therefore, focusing efforts on efficient acceleration using short pulses opens new potentials to realize high gradients, which in turn leads to low-emittance bunches and compact devices.

Generally, there is a \emph{conceptual gap} between standard accelerator technology and ultrafast science.
Microwave and millimeter-wave technology, used in conventional accelerators, are very well developed for producing continuous wave (CW) radiation.
Therefore, accelerators are mostly designed with narrowband excitations.
Examples are the widely used cascaded cavities which operate based on a resonant behavior and traveling wave accelerators, in which fields of a guided mode are employed for acceleration \cite{Wangler2008,Wiedemann2013}.
Hence, direct usage of a standard accelerator geometry excited by a short pulse laser incurs wasting a large portion of input energy.
The goal in this study is to introduce novel structures that aim to accelerate particles from rest using short pulse excitation, which we like to call \emph{single-cycle ultrafast electron guns}.

The last decade has witnessed extensive efforts on acceleration of electrons using optical pulses \cite{Peralta2013,England2014,Breuer2013,tajima1979,Leemans2006,wang2013,leemans2014,steinke2016}.
However, the acceleration schemes based on optical pulses suffer from the difficulties caused by the short optical wavelengths.
Some examples are emittance growth of the electron beam, increased energy spread in the bunch, and challenging timing synchronization for optical acceleration.
Research in THz pulse generation using optical rectification has led to single-cycle pulses \cite{Hoffmann2011,Fulop2011,Huang2013}.
The achieved performance in this process has reached percent level optical to THz conversion efficiency \cite{Huang2013,Schneider2014}.
Considering that picosecond lasers are necessary for single-cycle THz generation, which have been developed to much higher average power and pulse energies than 100\,fs type of lasers, THz acceleration using single-cycle pulses has become a viable option.
The first sub-keV devices are already realized and the predictions are evidenced based on experimental results \cite{ronny2016}.
Nonetheless, this scheme similarly demands broadband devices which function based on short pulse excitations.

This paper presents structures for accelerating particles using single-cycle THz pulses.
The considered temporal profile of the excitation is a single-cycle pulse described by $f(t)=A_0 \exp (-2 \ln2 (t-t_0 )^2 / \tau^2) \cos(\omega(t-t_0)+\phi_0 )$,
where $A_0=A_0(x)$ and $t_0=\pm x/c$ stand for the position dependent amplitude and phase, respectively and $\phi_0$ is the carrier envelope phase of the signal.
$\omega=2 \pi f_0$ denotes the angular frequency of the signal and $\tau=1/f_0$ is the pulse duration of the single-cycle pulse.
Note that the above solution is an approximate solution for a single-cycle pulse and suffers from the inaccuracy of containing a non-zero DC component.
However, the error is 0.001 of the peak-field which is not important for the purpose of our study.
Although we have considered the illustrated single-cycle pulse, the principle also works for few-cycle pulses, at the expense of additional energy.
We present structures which are useful in two different regimes of (i) low energy and (ii) high energy THz beams.

Detailed numerical simulations of the introduced structures play a central role in the presented research.
For this purpose, a DGTD/PIC code is employed, which captures all the involved field diffraction effects through the 3D full-vector time-domain solution of the Maxwell's equations using a Discontinuous Galerkin Time Domain (DGTD) method and computes the electron trajectories using a particle in cell (PIC) algorithm.
All the bunch evolution calculations in this study are carried out with the consideration of space-charge effects which is simulated using a point-to-point algorithm.
For more details on the implemented algorithm, the reader is referred to \cite{Fallahi2014}.
For initialization of macro-particles in the guns, we have used the ASTRA photoemission model \cite{Flottmann2011,Dowell2009}.
Note that ASTRA does not simulate the particle acceleration within transient fields and is merely used for bunch generation.

\emph{Low-energy Single-Cycle Ultrafast Electron Guns---} Based on the recent demonstration of 1\% level optical to THz conversion efficiencies \cite{Huang2011}, 2-mJ level slightly sub-ps pulses can safely generate 20-$\mu$J level single-cycle THz pulses typically at 300\,GHz central frequency.
If this beam is focused down to the diffraction limit, the total electric field at the focus with $2\lambda$ spot size ($\lambda$ is the central wavelength) is about 50\,MV/m.
In this field, initially at rest electrons are able to move maximally $\delta x= eE\tau/m\omega \simeq 7.5\,\mu\text{m}$, being 2-3 orders of magnitude smaller than the THz wavelength.
Consequently, the electrons are affected by both accelerating and decelerating cycles, leading to an inefficient acceleration process.
To acquire an efficient acceleration scheme, two goals must be achieved: 1) The accelerating field should be enhanced in order to lengthen the amplitude of electron vibration, and 2) the electron should leave the pulse before the decelerating cycle begins.

In \cite{ronny2016}, a simple planar device that pursues the above goals, is presented and experimented.
The planar gun consist of two metallic plates forming a structure like a 2D horn receiver antenna to focus the incoming linearly polarized THz beam below the diffraction limit.
The photocathode laser (usually a UV laser synchronized with the THz pulse) releases an electron bunch from the cathode surface, when the accelerating field of the THz pulse arrives at the injection point.
The electrons are then accelerated by the incoming THz beam and leave the acceleration region before the arrival of the decelerating cycle.
The study demonstrated acceleration of electrons with energies up to 0.8\,keV \cite{ronny2016}.

\begin{figure}
\includegraphics[width=6.0in]{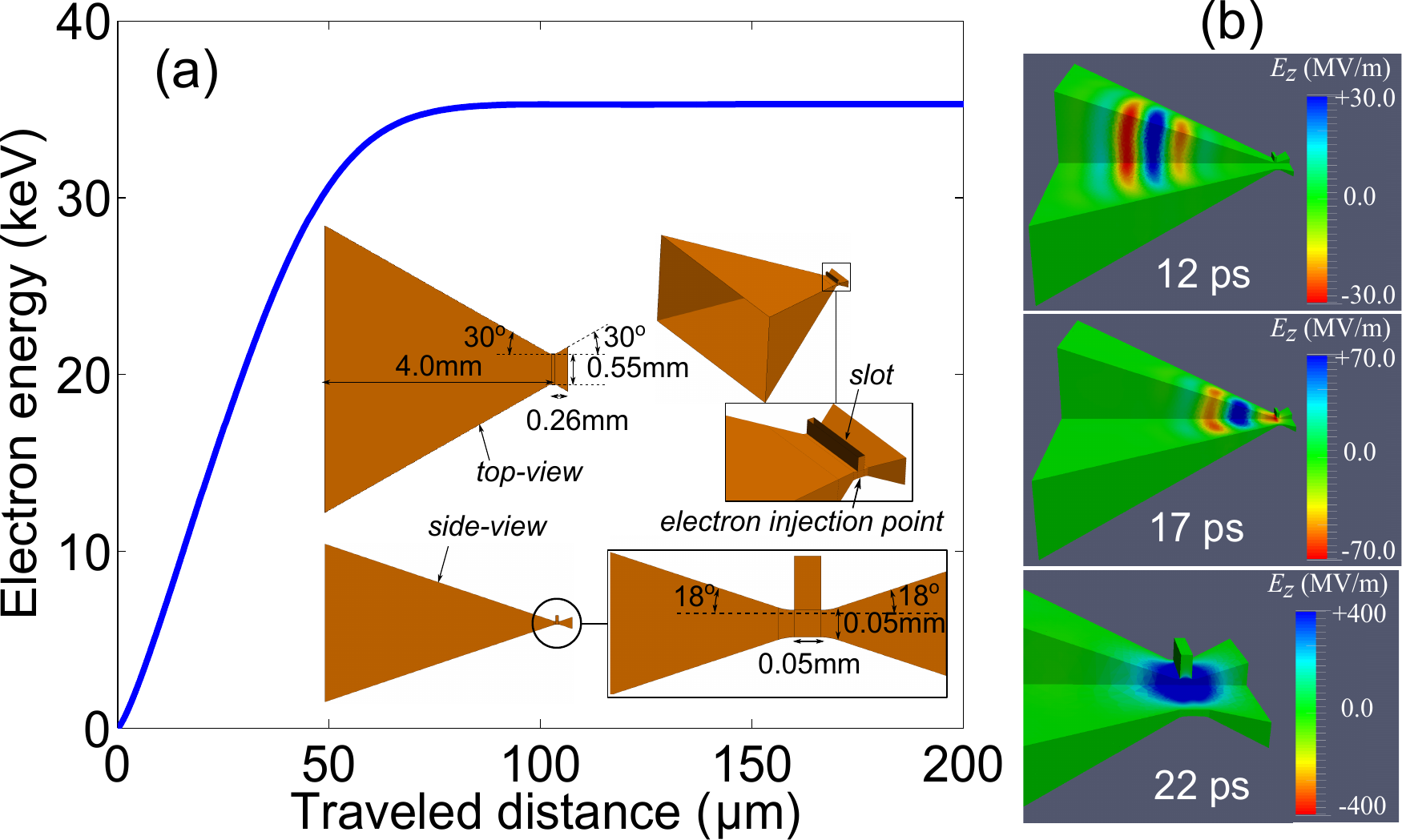}
\caption{(a) Schematic illustration of the low energy ultrafast electron gun and the energy of the accelerated electron versus the traveled distance. (b) Snapshots of ($E_z$) profile over two half-space cuts of the gun. Note the change in the color map scaling for different snapshots.}
\label{lowEGunFinal}
\end{figure}

There exist several techniques to enhance the efficiency of the concept of planar devices:
(1) In a planar device, the focusing of the THz beam is carried out only in the vertical plane, i.e. the E-plane. The same focusing can also be introduced in the H-plane to further enhance the accelerating field.
(2) The focusing in the H-plane introduces cut-off frequencies to the wave propagation. Therefore, the length of the injection region should be reduced to a fraction of the wavelength to enable the \emph{tunneling} of the accelerating field into the acceleration point.
(3) Adding a reflector at the receiving side of the structure with $\lambda/4$  distance from the electron injection point, causes the preceding decelerating half cycle to be inverted and added to the accelerating cycle upon reflection. Therefore, the total acceleration gradient at the injection point is enhanced.
(4) The same cut-off frequency effect holds also for the receiving side of the structure. Consequently, structuring the right side of the gun similar to the left side enhances the tunneling and thereby increases the acceleration gradient.
By taking the above considerations into account, an ultrafast electron gun driven by a low-energy single-cycle THz pulse is designed as shown in Fig.\,\ref{lowEGunFinal}a.
The energy of an electron at rest, injected at the instant with vertical field $E_z=50\,\text{MV/m}$, in terms of travel distance as well as snapshots of the accelerating field profile in the device are shown in Fig.\,\ref{lowEGunFinal}.
The simulations evidence an enhancement of the acceleration gradient by a factor of 15, leading to a peak acceleration field of $782\,\text{MV/m}$.
The final energy of the electron leaving the gun is $35.3\,\text{keV}$, being ideal for electron diffraction imaging.
\begin{figure}
\includegraphics[width=6.0in]{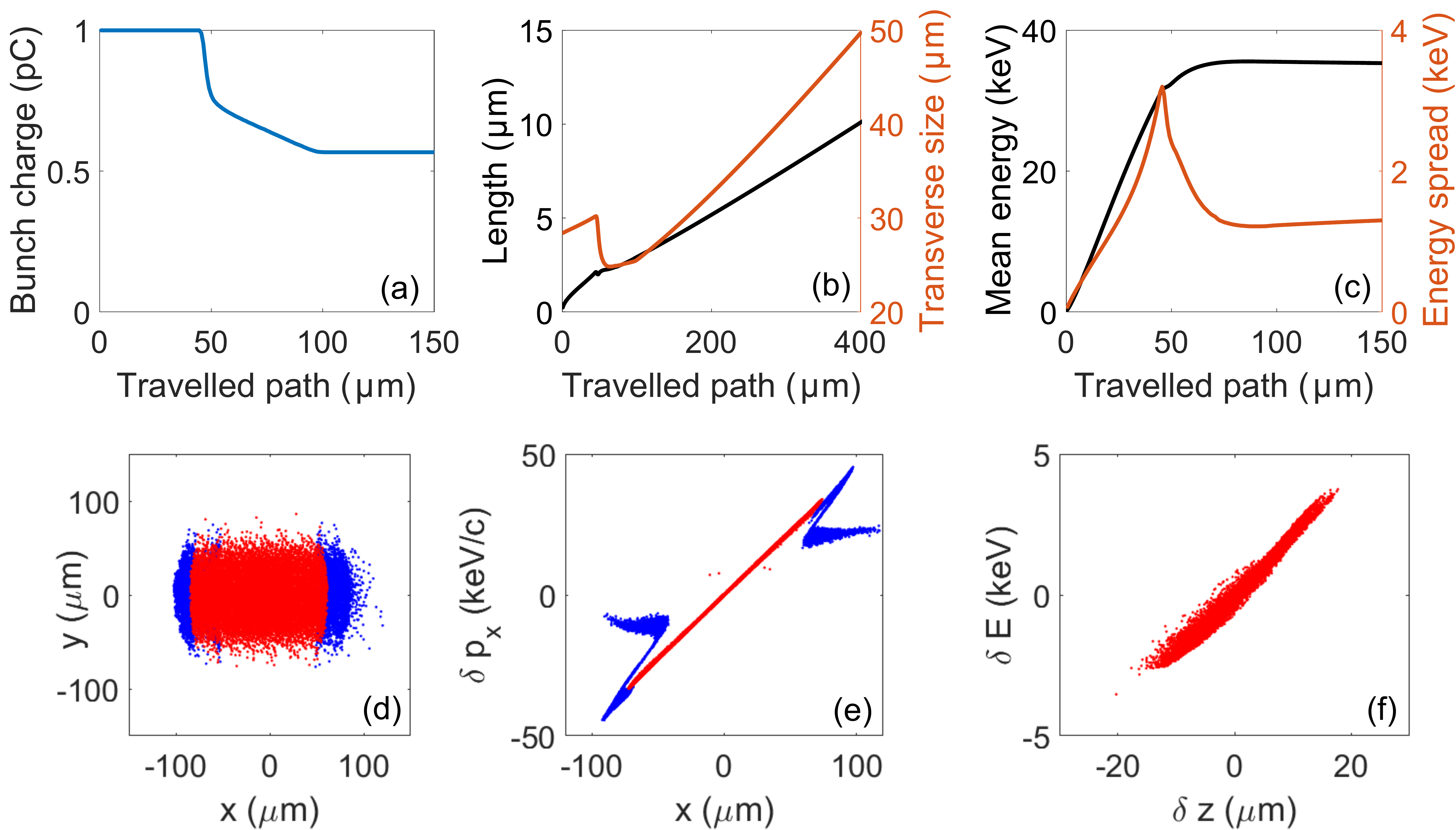}
\caption{Acceleration of 1\,pC photoemitted bunch in the high energy THz gun: (a) bunch charge, (b) size and (c) energy in terms of travelled path and (d) shape, (e) transverse phase space, and (f) longitudinal phase-space 100\,$\mu$m from gun exit are depicted (red dots: output particles; blue dots: trapped particles).}
\label{lowEGunFinalAcceleration}
\end{figure}

For the study of bunch evolution, we assume a copper cathode excited by a UV laser pulse with pulse duration equal to 40\,fs and spot size diameter 40\,$\mu$m.
The UV laser energy is chosen such that 1\,pC of charge is released which is modeled by 20'000 macro-particles.
The evolution of the bunch properties as well as snapshots of the bunch profile are shown in Fig.\,\ref{lowEGunFinalAcceleration}.
The mean energy of the bunch increases to 35\,keV with an energy spread of about 3\%, which happens due to the large spot size of the injected bunch compared to the THz wavelength (1\,mm).
It is observed that due to the collisions of the electrons with the metallic boundaries due to the transverse momentum of electrons (Fig.\,\ref{lowEGunFinalAcceleration}e), only 57\% of the photo-emitted electrons are extracted out of the gun.
This effect shows the limitation on the bunch size and correspondingly the amount of charge which can be accelerated with \emph{good} quality using the proposed THz gun.
Our simulations show that placing the introduced gun within a longitudinal magnetic field, i.e. producing the so-called magnetized beam, enhances the aptitude of the gun in terms of accelerated bunch charge.
For this purpose, structures producing 1\,T-level magnetic fields are required.
The final normalized emittances of the bunch are $(e_{n,x},e_{n,y},e_{n,z})$=(0.02,0.06,0.013)\,mm\,mrad and the output bunch length is about 95\,fs.

\begin{figure}
\includegraphics[width=6.0in]{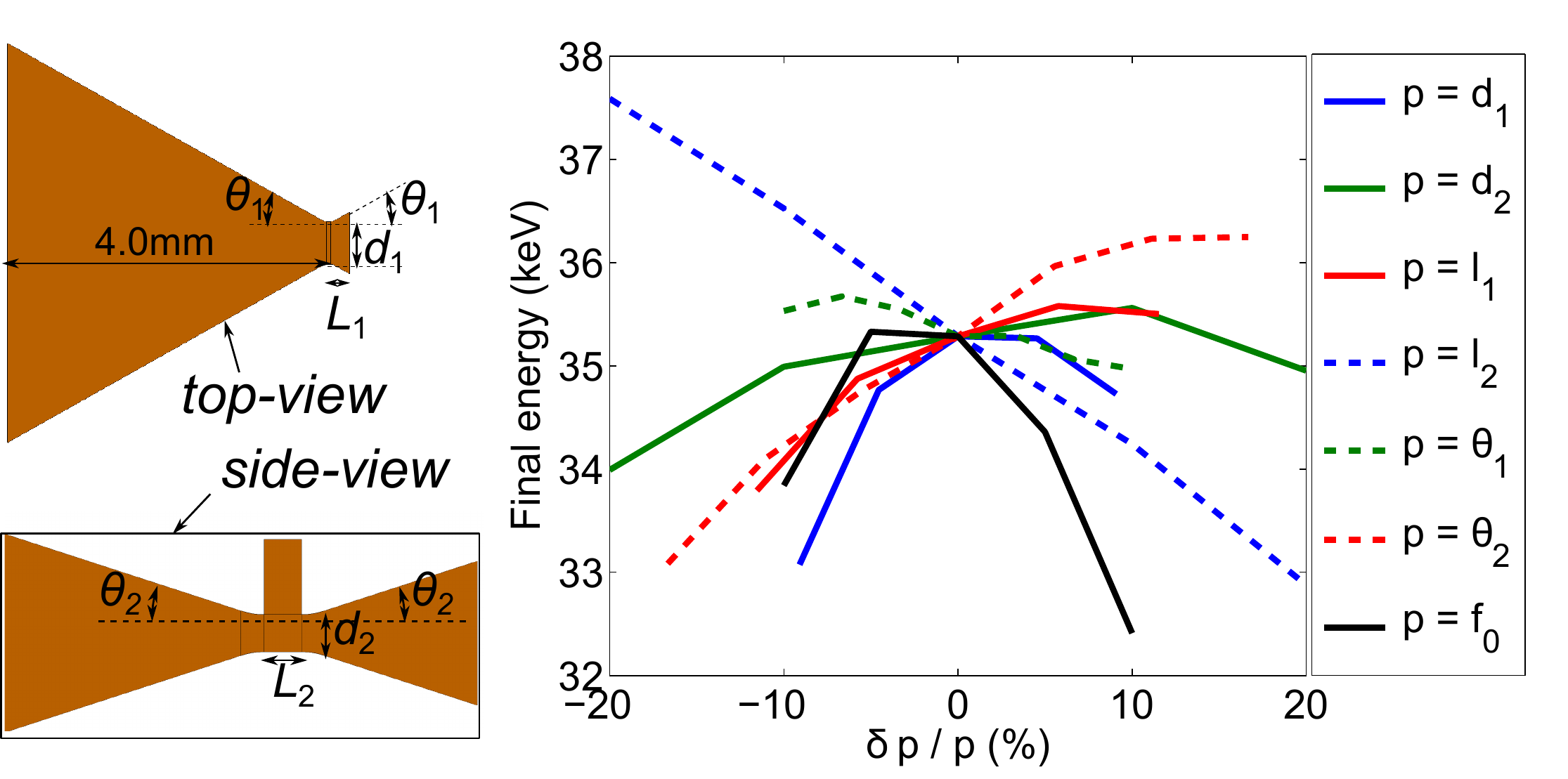}
\caption{Results of the sensitivity study for the low-energy ultrafast gun. The left figure introduces the studied parameters. }
\label{lowEGunSensitivity}
\end{figure}

In addition to the bunch acceleration study, a sensitivity analysis of the introduced gun is also of utmost importance.
For this purpose, we change each of the values independently by $\pm$10\% from the optimum design, study the THz propagation, inject electrons at the instant when the accelerating field is 50\,MV/m, and evaluate the acceleration performance.
The results of this analysis evidence maximum 3\% change in the final energy as the sensitivity of the device to implementation errors (Fig.\,\ref{lowEGunSensitivity}).
This is a very promising sensitivity compared to conventional electron guns, which has its origin in the broadband operation of the device.

\emph{High-energy Single-Cycle Ultrafast Electron Guns---} In the above designs, it was observed that optimum focusing of the THz beam with only 20\,$\mu$J energy leads to peak fields as large as 0.8\,GV/m on the electron emitter, which is close to the damage threshold of copper and other metals \cite{TantawiTalk}.
This means that increasing the energy of the input THz beam to achieve higher acceleration rates is not realistic.
However, today's THz generation technology has realized higher THz energy levels \cite{Schneider2014}.
Consequently, an issue is illuminated; how can one achieve efficient acceleration using high energy short pulses without surpassing the damage threshold?
Here, we try to answer this question by introducing structures which operate based on single-cycle THz beams with around 2\,mJ energy at 300\,GHz central frequency.

For this purpose, two important points must be taken into account:
(i) The electron may gain relativistic energy, which intensifies the effect of the transverse magnetic field of the THz pulse. This effect causes a push from the THz pulse along its propagation direction and leads to a curved trajectory for the electron motion.
(ii) A high-energy THz beam should not be focused to small spot-sizes to avoid dark current excitation.
As a consequence, to achieve an efficient acceleration, matching the phase front of the THz beam with the electron trajectory is essential.

\begin{figure}
\includegraphics[width=6.0in]{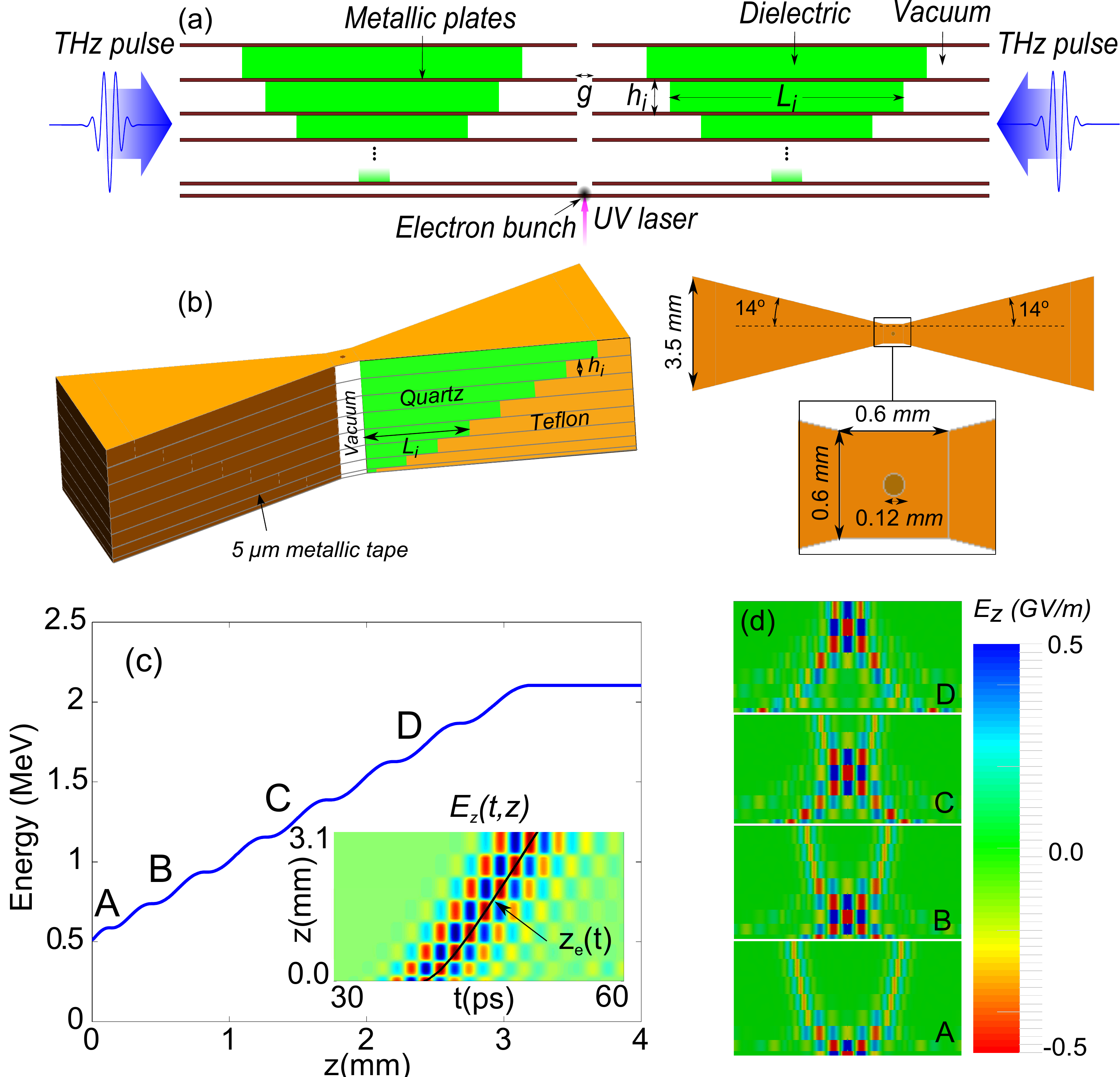}
\caption{(a) Schematic illustration of the high-energy ultrafast gun concept, (b) schematic illustration of the optimized ultrafast electron gun, (c) electron energy $E_e(z)$ and accelerating field $E_z(z,t)$ and (d) Snapshots of the field profile.}
\label{highEGunConceptAcceleration}
\end{figure}

The configuration illustrated in Fig.\,\ref{highEGunConceptAcceleration}a is a 2D presentation of the concept devised to solve the above two problems.
First, two linearly polarized THz beams are symmetrically coupled into the multilayer structure in order to cancel out the magnetic field where they overlap.
Second, the phase front of the THz beam is divided into several parts, which are isolated from each other using thin metallic surfaces.
In each layer, dielectric inclusions are added to delay the arrival time of the pulse to the acceleration region.
By properly designing the filling factor for each dielectric and the thickness of each layer, continuous acceleration of electrons from rest throughout the whole phase front can be achieved. %

As learned from the study on low energy guns, focusing the incoming excitation in the transverse plane enhances the acceleration efficiency.
Furthermore, to avoid suspended thin metallic films in vacuum (Fig.\,\ref{highEGunConceptAcceleration}a), we consider quartz ($\epsilon_r=4.41$) and teflon ($\epsilon_r=2.13$) for delaying the arrival time.
The structure shown in Fig.\,\ref{highEGunConceptAcceleration} is the ultrafast THz gun designed for operation based on two single-cycle THz Gaussian beams with each 1.1\,mJ energy and central frequency 300\,GHz.
Without loss of generality, the beam profile is flat top along the acceleration axis and Gaussian in the transverse direction with 2\,mm spot size.
For a completely Gaussian beam, individual couplers should be designed to couple the beam energy into the gun input (see supplementary material).
We assume that two linearly polarized plane waves with the aforementioned temporal signature and peak field $0.5\,$GV/m illuminate the multilayer gun from both sides. %
An eight layer configuration is designed for the considered excitation with the thickness of each layer $h_i=\{0.1,0.27,0.35,0.40,0.42,0.43,0.44,0.45\}$\,mm, and the length of the quartz inclusions $L_i=\{0.2,0.865,1.55,2.25,2.95,3.7,4.4,5.1\}$\,mm.
The size of the acceleration channel is considered to be $g=120\,\mu$m.
The simulation results shown in Fig.\,\ref{highEGunConceptAcceleration}c demonstrate acceleration of an electron from rest to 2.1\,MeV.
Similar to the previous cases, the electron is released at the point with $E_z=50$\,MV/m.
The realization of phase front matching with the electron motion is shown by snapshots of the field profile in Fig.\,\ref{highEGunConceptAcceleration}d and the subfigure in Fig.\,\ref{highEGunConceptAcceleration}c showing the accelerating field $E_z(t,z)$ superposed on the particle position $z_e(t)$.
The small average momentum of the particles in the bottom layers causes small travelling distances within one half-cycle.
Therefore, the thicknesses of the layers need to be smaller than the top layers to achieve the required synchronism.
This effect leads to the observed gradual increase in the energy gain in different layers (Fig.\,\ref{highEGunConceptAcceleration}b).

\begin{figure}
\includegraphics[width=6.0in]{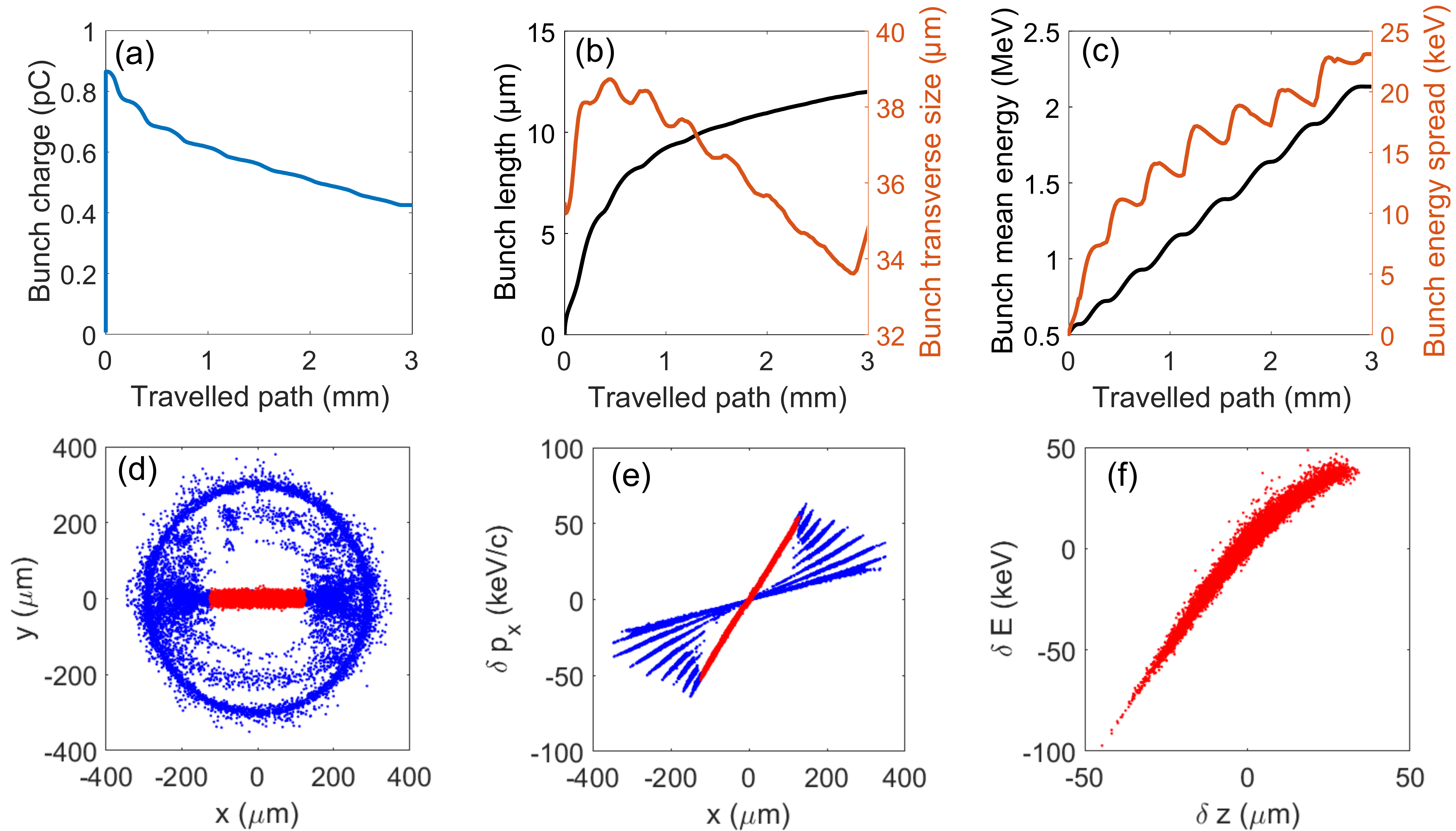}
\caption{Acceleration of 1\,pC photoemitted bunch in the high energy THz gun: (a) bunch charge, (b) size and (c) energy in terms of travelled path and (d) shape, (e) transverse phase space, and (f) longitudinal phase-space 3\,mm from gun exit are depicted (red dots: output particles; blue dots: trapped particles). }
\label{highEGunFinalAcceleration}
\end{figure}

By again initializing a photo-emitted electron bunch, the bunch evolution in the proposed gun is simulated.
We assume that a 40\,fs UV laser pulse generates 1\,pC charge over a 60\,$\mu$m spot size, which is modeled using 20'000 macro-particles.
The simulation results (Fig.\,\ref{highEGunFinalAcceleration}) demonstrate acceleration of 40\% of the bunch to 2.1\,MeV with 1.1\% energy spread, output  bunch length of 45\,fs, and normalized emittances equal to $(e_{n,x},e_{n,y},e_{n,z})$=(0.13,0.13,0.09)\,mm\,mrad.
The main reason for the particle loss is the deflection of the electron trajectories out of the considered vertical path and collision with the metallic layers, which could again be mitigated by focusing coils (see the supplementary material).

\begin{figure}
\includegraphics[width=6.0in]{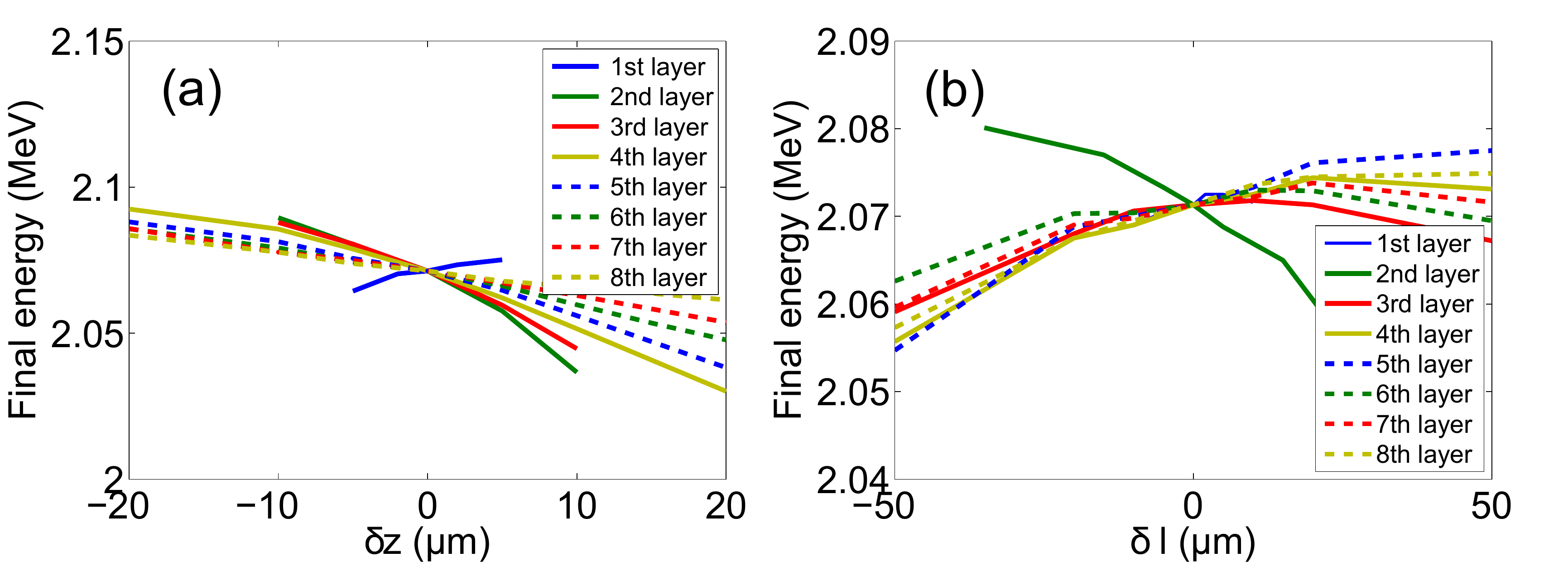}
\caption{Sensitivity of the ultrafast high-energy gun to variations in (a) layer thicknesses and (b) length of quartz inclusions.}
\label{sensitivity}
\end{figure}

Our detailed investigations of the introduced high energy ultrafast gun showed several advantages of such a scheme compared to conventional cascaded or travelling wave cavities.
First, due to the inherent operation of the device with broadband excitations, the sensitivity of the gun outcome to the design parameters is minimal.
This is illustrated in Fig.\,\ref{sensitivity} for the variations in dielectric lengths and layer thicknesses.
It is observed that even 10\% change in the designed parameter is tolerated by the device.
Second, the pulse heating due to the magnetic field is not only minimized by the single cycle operation, but also the magnetic field is canceled at the electron acceleration region.

In conclusion, we have presented various structures that can serve as miniaturized electron guns excited by single-cycle THz pulses.
Devices for both low energy ($\mu$J level) and high energy (mJ level) THz pulses are proposed.
The maximum normal electric field on the surface was allowed to be as high as 0.8\,GV/m, although the scaling equations predict much higher thresholds for extremely short pulse excitations.
This leads to potential additional improvements in terms of acceleration gradient and justifies the use of ultrafast structures to achieve compact accelerators.
The presented ultrafast electron guns are promising devices to realize short bunches for applications in electron diffractive imaging, microscopy and compact radiation sources.

\emph{Acknowledgement---} This work is supported by the European Research Council (ERC) under grant 609920. A. Yahaghi acknowledges support by the Alexander von Humboldt foundation.

%


\end{document}